 \documentclass[a4paper,useAMS,usenatbib]{mnras}

\usepackage[T1]{fontenc}
\usepackage{ae,aecompl}

\usepackage{graphicx}	
\usepackage{amsmath}	
\usepackage{amssymb}	

\usepackage{longtable}

\usepackage{times}
\usepackage{hyperref}
\usepackage{color}
\usepackage{xcolor}
\usepackage{multicol}
\usepackage{booktabs}
\usepackage{caption}
\usepackage{txfonts}
\usepackage{ulem,color}
\usepackage{url}

\newcommand{\teff}{$T_\textrm{eff}$}
\newcommand{\logg}{$\log g$}
\newcommand{\feh}{[Fe/H]}
\newcommand{\luminosity}{$\log (\rm{L/L_{\odot}})$}
\newcommand{\kms}{km\,s$^{-1}$}
\newcommand\Tstrut{\rule{0pt}{2.2ex}}         

\title[High-resolution Spectroscopy and Spectropolarimetry]{High-resolution Spectroscopy and Spectropolarimetry of Selected $\delta$-Sct Pulsating Variables}

\author[Joshi et al.]
{Santosh Joshi$^1$\thanks{Email:santosh@aries.res.in}, Eugene Semenko$^{2}$, A. Moiseeva$^2$, Kaushal Sharma$^3$, Y. C. Joshi$^1$, 
\newauthor{M. Sachkov$^{4,5}$, Harinder P. Singh$^3$ and Yerra Bharat Kumar$^6$} \\\\
$^1$ Aryabhatta Research Institute of Observational Sciences, Manora Peak, Nainital- 263002, India\\
$^2$ Special Astrophysical Observatory of the Russian Academy of Sciences, Nizhnii Arkhyz,    Karachai-Cherkessian Republic - 369167, Russia\\
$^3$ Department of Physics and Astrophysics, University of Delhi, Delhi - 110007, India\\
$^4$ Institute of Astronomy, Russian Academy of Sciences, Moscow- 119017, Russia\\
$^5$ Institute of Applied Technical and Economic Research and Expert Assessment, RUDN University, Moscow - 117198, Russia\\
$^6$ Key Laboratory of Optical Astronomy, National Astronomical Observatories, Chinese Academy of Sciences, 100012 Beijing, PR China
}
\date{Released 2016 Xxxxx XX}

\pubyear{2016}

\begin{document}
\label{firstpage}
\pagerange{\pageref{firstpage}--\pageref{lastpage}}
\maketitle

\newcommand{\BibTeX}{{\sc Bib}\TeX}
\setlength{\bibsep}{0.1pt}


\begin{abstract}
	
The combination of photometry, spectroscopy and spectropolarimetry of the chemically peculiar stars often aims to study the complex physical phenomena such as stellar pulsation, chemical inhomogeneity, magnetic field and their interplay with stellar atmosphere and circumstellar environment. The prime objective of the present study is to determine the atmospheric parameters of a set of Am stars to understand their evolutionary status. Atmospheric abundances and basic parameters are determined using full spectrum fitting technique by comparing the  high-resolution spectra to the synthetic spectra. To know the evolutionary status we derive the effective temperature and luminosity from different methods and compare them with the literature. The location of these stars in the H-R diagram demonstrate that all the sample stars are evolved from the Zero-Age-Main-Sequence towards Terminal-Age-Main-Sequence and occupy the region of $\delta$ Sct instability strip. The abundance analysis shows that the light elements e.g. Ca and Sc are underabundant while iron peak elements such as Ba, Ce  etc. are overabundant and these chemical properties are typical for Am stars. The results obtained from the spectropolarimetric analysis shows that the longitudinal magnetic fields in all the studied stars are negligible that gives further support their Am class of peculiarity.

\end{abstract}


\section{Introduction}
\label{section:intro}
	
A fraction of early-B to late-F Main-Sequence (MS) stars characterized by abnormal strengths of spectral lines forms a group of chemically peculiar (CP) stars. These stars show overabundance of iron peak, rare-earth elements and underabundance of light elements such as Ca and Sc in their photosphere \citep{1988JBAA...98..116J}. Possibly, modeling of these stars is most challenging due to the presence of convection, magnetic field, segregation and stratification of the chemical elements in the same regions of the atmosphere. The chemical peculiarities in these stars are explained by the microscopic diffusion resulting from the competition between gravitational settling towards the center and uplift to the surface by the radiation field \citep{2012MNRAS.425.2715S}. In the presence of the significant magnetic field, the diffusion process is guided by the magnetic force acting on different ions \citep{1992A&A...258..449B}. As a result, inhomogeneities of chemical elements are frequently seen in the form of spots and cloud-like structures in their surface. The observed peculiar abundance of chemical species is related to the stellar photosphere and does not reflect the chemical composition of the entire star.

According to \citet{1974ARA&A..12..257P}, the CP stars are basically divided into four groups: metallic line stars (CP1), magnetic Ap stars (CP2), HgMn stars (CP3), and He-weak stars (CP4). 
The CP1 and CP3 stars are considered as non-magnetic and CP2 and CP4 stars exhibit globally organized magnetic fields with typical strength from $\sim$ 300 G \citep{2007A&A...475.1053A} up to 30 kG \citep{2010MNRAS.402.1883E}. \citet{1984A&A...138..493M} proposed an extension of Preston's CP scheme with CP5 class contains He-weak magnetic stars while CP6, the non-magnetic He-weak stars.

Using the state-of-art instruments such as NARVAL and ESPaDOnS, the discovery of ultra-weak magnetic field
($<$ 1 G) in Am stars \citep{2011A&A...532L..13P, 2016MNRAS.459L..81B} and $\delta$ Sct type stars \citep{2015MNRAS.454L..86N} are reported through highly precise spectropolarimetric observations using stable spectropolarimeters. The detection of low-order magnetic field in non-magnetic stars (e.g. Am stars) allows us to study the interaction of weak magnetic field with various  physical processes.

\begin{table*}
\caption{Basic parameters of sample stars taken from literature. The distance d is calculated using parallax $\pi$ extracted from GAIA DR1 catalogue \citep{Gaia_Collaboration2016}. }
\label{table:samples}
\scriptsize
\begin{center}
\begin{tabular}{lcccccccccccccr}
\hline
					&& \\
					Star & $\alpha$ & $\delta$ &$m_v$ & $\pi$ & d& $\rm{P_1}$&$\rm{P_2}$&  $b-y$ & $m_1$ & $c_1$ & $\beta$ & $B-V$ & $V-I$& Sp. Type$^a$\\
					&  &       &     &   (mas)   &  (pc)   & (min) & (min) & & & & &\\
					&& \\
					\hline
					&& \\
					HD\,13038& 02 09 30 & +57 57 38&   8.56 &   --  & --   & 28.7 &  34.0 & 0.105 & 0.220 & 0.848 & 2.856 & -- & -- &  --  A4       \\
					HD\,13079 &  02 09 02 & +39 35 32  &  8.79 & 5.84$\pm$0.25 & 171 & 73.2 &  -- & 0.203 & 0.211 & 0.672 & 2.759 & -- & --  & F0--      \\
					HD\,25515 & 04 05 36 & +50 45 33  &   8.66 & 5.21$\pm$0.32 & 192 &166.8 &  --  &  0.262 & 0.177 & 0.745 & 2.706 &0.408 & $0.470$ & F0--dD   \\
					HD\,98851 & 11 22 51 & +31 49 41   &   7.40 & 6.51$\pm$0.29 & 154  & 81.0 & 162.0 &  0.199 & 0.222 & 0.766 &   --   &0.335 & $0.390$ & F1--F3 Sr \\
					HD\,102480 & 11 47 52 & +53 00 54 &  8.45 & 3.62$\pm$0.27 & 276  &156.0 &  84.0 &  0.211 & 0.204 & 0.732 &   --   & 0.360 & $0.420$ & F2--F5 Sr \\
					HD\,113878 & 13 06 00 & +48 01 41  &   8.26 & 3.44$\pm$0.26 & 291  &138.6 &  --  &  0.219 & 0.257 & 0.729 & 2.745 & 0.371 & $0.430$ &  A9--F3 Sr \\
					HD\,118660 &  13 38 07 & +14 18 06 &  6.30 & 13.94$\pm$0.51 & 72 & 60.0 & 151.2 &  0.150 & 0.214 & 0.794 & 2.778 & 0.274 & $0.310$ & A6--F0    \\
					\\
					\hline
				\end{tabular}
		\end{center}
\begin{minipage}{\textwidth}%
Notes: $^a$ Spectral types are adopted from \cite{2009A&A...498..961R}, where dash separates the spectral type computed from the K line and from the metallic lines. $\delta$  Del-type stars are classified as Am, and marked with `dD'. Spectral types are followed by elements with abnormal abundances.
\end{minipage}
		\end{table*}

	The normal A-type stars generally exhibit nearly solar elemental abundances in their photosphere except some elements heavier than iron show over and/or underabundance  within 0.4 dex relative to the Sun \citep{2007BaltA..16..183A}. In comparison to normal A-type stars, the Am stars show deficiency of Sc and Ca, enhancements of the iron peak elements and overabundance of Sr, Y, Zr, and Ba~\citep{2009A&A...503..945F}. The effective temperature of Am stars lie in the range of 7000--9000 \,K making them coolest CP stars on the MS.   Most of the Am stars are found in binary system with orbital periods between 2 to 10 days ($v \sin i$ $<$ 120 km\,s$^{-1}$) where the low-rotational velocity is reduced by tidal interaction \citep{2008JKAS...41...83T}. One important question in the context of Am stars is the excitation of pulsations driven by $\kappa$-mechanism operating in the He~\textsc{ii} ionization region \citep{2000ASPC..210..215P}. The normal A-type stars are usually rapid rotators where helium is mixed-up by turbulence induced meridional circulation, thus one can expect the excitation of the pulsations in the presence of He in He~\textsc{ii} ionization zone \citep{1979ApJ...231..798C, 2000A&A...360..603T}. In contrast to the normal stars, Am stars rotate slowly, hence the helium from the He~\textsc{ii} ionization zone is almost depleted due to gravitational settling. As a result the deficiency of He in He~\textsc{ii} ionization zone prohibits the operation of $\kappa$-mechanism, hence  Am stars should not pulsate. However, this scenario  has been changed thanks to highly precise  data obtained from ground and space based survey telescopes \citep{smalley2011, 2011MNRAS.414..792B}. Observations from these facilities clearly demonstrate that many Am stars show multi-periodic $\delta$ Sct  pulsational variability \citep{2014A&A...564A..69S}, hence they are considered as potential candidates for asteroseismology \citep{2015JApA...36...33J}. Briefly, the $\delta$ Sct stars are pulsating stars of mass range between 1.5 $M_{\odot}$ to 2.5 $M_{\odot}$ and they are confined in the region of H-R diagram where MS crosses the lower extension of the Cepheid instability strip \citep{2000ASPC..210.....B, 2010aste.book.....A}. The pulsation periods of the $\delta$ Sct stars are ranging from $\sim $0.5 to 7 hr and the pulsations in these stars are driven by $\kappa$-mechanism \citep{2014A&A...564A..69S}. The pulsating Am stars show variability similar to the $\delta$ Sct type but with lower amplitude to their normal counterpart \citep{kurtz1989}.

The Nainital-Cape survey is an international project commenced in 1999 between Aryabhatta Research Institute of Observational Sciences (ARIES) {\it Nainital}, India  and South African Astronomical Observatory (SAAO) {\it Cape Town} South Africa with an aim to search for and study the pulsations in Ap and Am stars using the high-speed photometry~\citep{2001A&A...371.1048M, 2006A&A...455..303J,joshi2009,2016A&A...590A.116J}.

	\begin{table}
		\caption{Observational log of high-resolution spectroscopic observation of sample stars.}
		\label{table:spectro}
		\begin{center}
			\begin{tabular}{lccr}  
				\hline
				Star & HJD(2450000+) & Sp. range  & SNR \\
				HD   & (day)         & (\AA)      &     \\
				\hline
				13038  & 5025.359 & 4840--6240 & 150 \\
				       & 6236.385 & 4385--6670 & 110 \\
				13079  & 6587.452 & 4400--6780 & 100 \\
				25515  & 6587.367 & 4400--6780 & 120 \\
				98851  & 5025.265 & 4840--6240 & 260 \\
				113878 & 5025.312 & 4840--6240 & 220 \\
				118660 & 6408.505 & 3950--6860 & 130 \\
				\hline
				\noalign{\smallskip}
			\end{tabular}
		\end{center}
	\end{table}

In addition to photometry, spectroscopic and spectropolarimetric study of the CP stars are equally important for the study of pulsational variability, determination of their atmospheric parameters, chemical abundances, magnetic field strength etc. For this, we set-up an Indo-Russian collaboration to study the CP stars using spectroscopic and spectropolarimetric  observations using the observational facilities available in India and Russia. As of now, we have performed  the combined photometric and spectroscopic analysis of three Am stars namely, HD\,98851 \& HD\,102480 \citep{2003MNRAS.344..431J} and HD\,207561 \citep{2012MNRAS.424.2002J}, and an Ap star HD\,103498 \citep{2010MNRAS.401.1299J}. The current  paper is fourth of this series presenting the detail spectroscopic and spectropolarimetric analysis of seven Am stars where $\delta$ Sct pulsating variability was discovered under the Nainital-Cape survey project. This manuscript is structured  as follows: In Sec.~\ref{section:observation} we have given the details of observation and data reduction techniques. Derived stellar parameters from photometry and spectroscopy are presented in Sec.~\ref{section:para}. We present the magnetic field measurements in Sec.~\ref{section:magnetic}. Abundance analysis of the target stars is discussed in Sec.~\ref{section:abundance}. The luminosity of studied samples obtained from different methods followed by their evolutionary status is described in Sec.~\ref{sec:hrd}. Detailed information on individual star and the comparison of their observed parameters to the literature is given in Sec.~\ref{sec:individual_stars}. Finally, the discussion and conclusion drawn from our study is presented in Sec.~\ref{section:conclusion}.

		\begin{figure}
			\centering
			\includegraphics[width=\columnwidth]{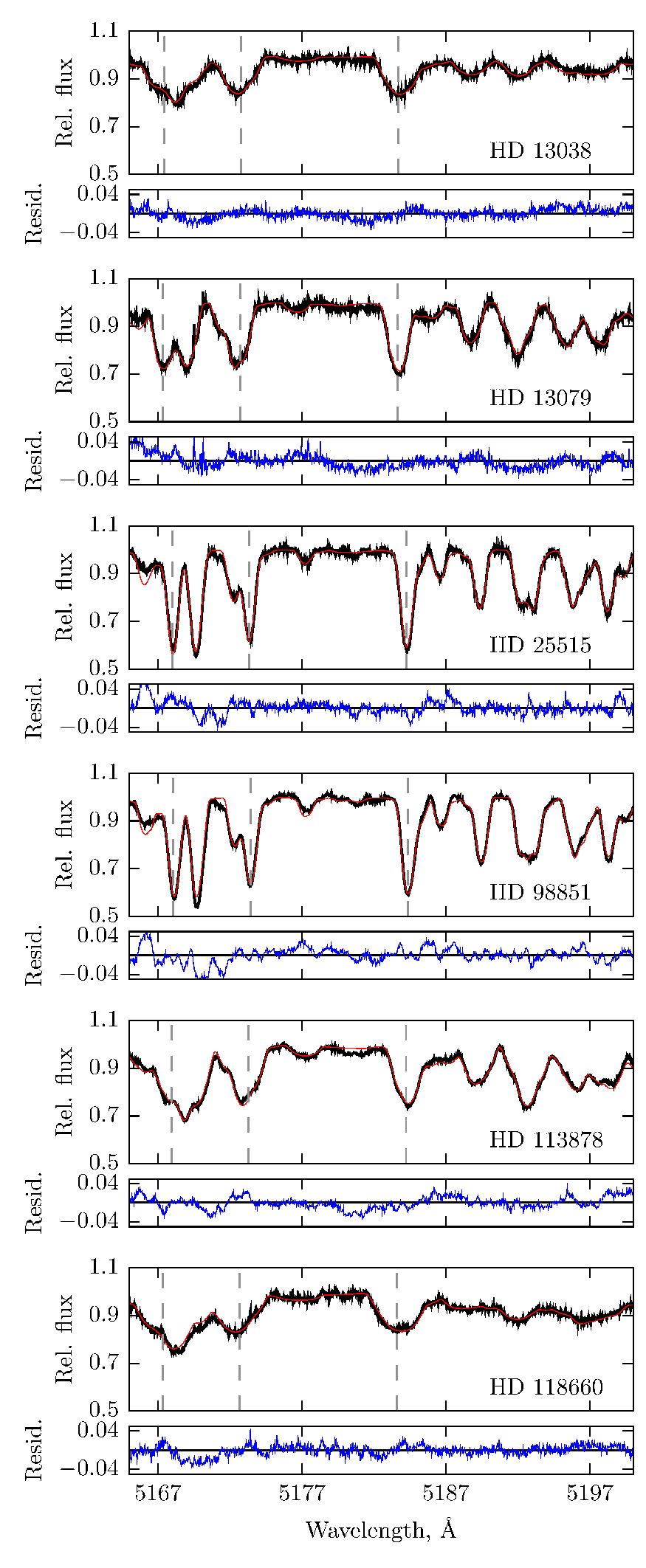}
			\caption{Examples of observed spectra of the sample stars shown by black line. Rotationally broaden spectral lines are fitted using synthetic spectra (red line). This wavelength region was used to derive the surface gravity \logg {} spectroscopically.  The lower panels show the residuals of the best fit.}
			\label{figure:nes}
		\end{figure}

	\section{Target Selection, Observation and Data Reduction}
	\label{section:observation}
	
Our sample consists of stars discovered as  $\delta$ Sct type pulsating variables under the Nainital-Cape survey project \citep{1999MNRAS.309..871M, 2001A&A...371.1048M, 2003MNRAS.344..431J, 2006A&A...455..303J, joshi2009, 2016A&A...590A.116J}. The  Am candidates along with their basic parameters are given in Table~\ref{table:samples}.

High-resolution spectroscopic observations of candidate stars were carried out to determine the basic physical parameters and their chemical abundances. For high-resolution spectroscopy, we used Nasmyth Echelle Spectrometer (NES) mounted on 6.0-m telescope BTA at Special Astrophysical Observatory (SAO) located at the North Causasus region of Russia. Two different CCD detectors were used during the period of our observations. Initially, a  CCD with $2048\times2048$ pixels was used for recording the spectra and later this detector was decommissioned in 2011. Afterwards, a  CCD detector of $2050\times4600$ pixels size with improved sensitivity in the UV region was used to register the spectra. The pixel size of both detectors is 13.5\,$\mu$, hence the mean resolving power (R$\sim$39,000) of NES spectrograph remained nearly same. During the observations, a set of bias and flat-field images were obtained for removing the pixel noise and correcting the pixel sensitivity, respectively. For the purpose of wavelength calibration a single Th-Ar arc spectrum was acquired just before or after the target star spectrum. Preliminary data reduction was performed uniformly using IDL-based software REDUCE~\citep{2002A&A...385.1095P}. The sequence of spectral data reduction consists of subtraction of averaged bias and scattered light, the flat-field correction and the wavelength calibration. Before further analysis,  all the spectra  were normalized to the continuum level using the {\sc continuum} task of IRAF\footnote{IRAF is distributed by the National Optical Astronomy Observatories, which is operated by the Association of Universities for Research in Astronomy, Inc., under contract to the National Science Foundation.}. In Table \ref{table:spectro}, we have given the details information on the high-resolution spectroscopic observations where signal to noise ratio (SNR) is calculated at central wavelength 5500 \AA. A portion of the \'{e}chelle spectra of the sample stars are shown in Fig. \ref{figure:nes}. The upper panels of this figure show fitting of observed spectra to the sythetic while lower ones show residuals of the best fit.

	Spectropolarimetric observations of sample stars were obtained using Main Stellar Spectrograph (MSS) mounted on 6.0-m telescope of SAO. MSS is a long slit spectrograph equipped with a circular polarization analyzer combined with double image slicer and a rotatable quarter-wave plate \citep{2004mast.conf..286C}. The quarter-wave plate is able to take polarized spectra at two fixed positions corresponding to the angles $-45^{\circ}$ and $+45^{\circ}$ relative to the principal axes of a birefringent crystal. The detector used on MSS is a $2050\times4600$ CCD of pixel size 13.5\,$\mu$. Polarized spectra from the MSS cover wavelength range of 530\,\AA\ in the 3$^\mathrm{rd}$ order with the average spectral resolution of R$\sim$14500. This instrument is capable of measuring the longitudinal magnetic fields in a stellar source as faint as $m_v$ $\sim$ 12. The typical accuracy of measurements in magnetic field is about 50 G depending upon  total number of  lines used for the measurements, the width of the spectral lines and strength of the magnetic field.

	Polarized spectra of sample stars were obtained in a standard way comprising a series of paired exposures in two orthogonal orientations of a retarder. The data reduction was performed using \textsc{zeeman} package developed within the European Southern Observatory-Munich Image Data Analysis System\footnote{\url{http://www.eso.org/sci/software/esomidas/}} \citep[\textsc{eso-midas,}][]{Banse1983,Warmels1992} for a specific format of CCD spectra taken with an image slicer~\citep{2006MNRAS.372.1804K}. Apart from the common steps in reduction of the long slit spectral data (e.g. bias subtraction, flat fielding and cosmic rays removal) \textsc{zeeman} uses creation of a mask with individual traces for slices and extraction of one dimensional polarized spectra to the individual arrays. Finally, the one dimensional spectra were normalized to the continuum level using the \textsc{continuum} task from IRAF. The observational details of the spectropolarimetric observations is given in Table \ref{table:spectropol} where SNR is calculated at central wavelength 4550 \AA. The intensity and circularly polarized spectra are shown in Fig. \ref{fig:polarim}. 

\begin{table}
		\caption{Detailed information on the spectropolarimetric observations of sample stars.}
		\label{table:spectropol}
		\begin{center}
			\begin{tabular}{lccr}  
				\hline
				Star & HJD(2450000+) & Sp. range  & SNR \\
				HD   &   (day)       & (\AA)      & \\  
				\hline
				13038  & 6644.408 & 4428--4983 &  250 \\
				13079  & 6644.383 & 4428--4983 &  200 \\
				25515  & 6644.490 & 4428--4983 &  250 \\
				98851  & 6383.463 & 4434--4982 &  370 \\
				102480 & 6644.533 & 4428--4983 &  270 \\
				113878 & 6381.497 & 4430--4985 &  180 \\
				           & 6383.486 & 4428--4984 &  250 \\
				118660 & 6383.521 & 4428--4984 &  350 \\
				           & 6383.538 & 4428--4984 &  330 \\
				\hline
			\end{tabular}
		\end{center}
	\end{table}

\section{Stellar Parameters} 
\label{section:para}

In order to determine the photospheric abundances of various elements from the observed spectra, one should have  precise value of atmospheric parameters e.g. effective temperature (\teff{}), surface gravity (\logg{}) etc. To derive these physical quantities, we used two traditional approaches, namely, photometric calibrations and high-resolution spectral analysis method. In the following sub-sections, we describe each method in a greater detail.
		
		\begin{table*}
\caption{The derived and compiled atmospheric parameters of the programme stars.}
\label{table:final_results}
\begin{center}
\scalebox{0.95}{
\begin{tabular}{lcccccl} 
\hline
Star        &   \teff{}         & \logg{}            & $v\sin i$     & $\xi_\mathrm{mic}$& \luminosity{}   & References    \\
            &   (K)             & (dex)              &  (\kms{})     &      (\kms{})     &  (dex)        &              \\\hline
HD\,13038   & 7960\,$\pm$\,200  & 3.80\,$\pm$\,0.15  &  87\,$\pm$\,5 &  1.8\,$\pm$\,1.1  & 1.51\,$\pm$\,0.19 &    1a        \Tstrut\\
            &   8090            &      -             &      -        &         -         &       -           &      1b       \\
            &   7890            &    4.34            &      -        &         -         &       -           &     1c       \\ 
            &   8147            &      -             &      -        &         -         &     0.76          &     2        \\
            &   8210            &      -             &      -        &         -         &     0.79          &     4        \\\hline
HD\,13079   & 7040\,$\pm$\,200  & 3.40$\,\pm$\,0.20  & 56\,$\pm$\,3  &  3.7\,$\pm$\,0.5  & 0.97\,$\pm$\,0.22 & 1a        \Tstrut\\
            &   7230            &      -             &      -        &         -         &       -           &      1b       \\
            &   7080            &    4.25            &      -        &         -         &       -           &     1c       \\ 
            &   7482            &      -             &      -        &         -         &     0.62          &      2        \\
            &   7270            &      -             &      -        &         -         &       -           &     3        \\
            &   7370            &      -             &      -        &         -         &     0.68          &      4        \\
            &   7200            &    4.00            &     45        &         2         &       -           &     7        \\
            &   7145            &      -             &      -        &         -         &     0.74          &      9        \\\hline
HD\,25515   & 6650\,$\pm$\,250  & 3.80$\,\pm$\,0.15  & 38\,$\pm$\,8  &  3.8\,$\pm$\,0.8  & 1.01\,$\pm$\,0.26 &     1a        \Tstrut\\
            &   6970            &      -             &      -        &         -         &       -           &     1b       \\
            &   6850            &    3.46            &      -        &         -         &       -           &    1c       \\
            &   6461            &      -             &      -        &         -         &       -           &     3        \\
            &   6823            &      -             &      -        &         -         & 0.97\,$\pm$\,0.34 &     6        \\\hline
HD\,98851   & 7000\,$\pm$\,200  & 3.65$\,\pm$\,0.15  & 40\,$\pm$\,5  &  3.7\,$\pm$\,0.3  & 1.43\,$\pm$\,0.21 &    1a        \Tstrut\\
            &   7300            &      -             &      -        &         -         &       -           &      1b       \\
            &   7150            &    3.63            &      -        &         -         &       -           &    1c       \\
            &   6999            &    3.50            &      -        &         -         &     1.67          &      2        \\
            &   6800            &      -             &      -        &         -         &       -           &     3        \\
            & 7000\,$\pm$\,250  & 3.50\,$\pm$\,0.50  &      -        &         -         & 1.50\,$\pm$\,0.10 &   5        \\\hline
HD\,102480  & 6720\,$\pm$\,250  & 2.90$\,\pm$\,0.20  & 40\,$\pm$\,5  &  3.5\,$\pm$\,0.5  & 1.41\,$\pm$\,0.26 &     1a        \Tstrut\\
            &   7180            &      -             &      -        &         -         &       -           &     1b       \\
            &   7040            &    3.90            &      -        &         -         &       -           &     1c       \\
            &   6966            &    3.11            &      -        &         -         &     2.18          &     2        \\
            &   6967            &      -             &      -        &         -         &       -           &      3        \\
            & 6750\,$\pm$\,250  & 3.00\,$\pm$\,0.50  &      -        &         -         & 1.40\,$\pm$\,0.20 &     5        \\
            &   7000            &    3.80            &      -        &         -         &       -           &    8        \\\hline
HD\,113878  & 7000\,$\pm$\,200  & 3.70$\,\pm$\,0.10  & 74\,$\pm$\,6  &  3.2\,$\pm$\,0.3  & 1.53\,$\pm$\,0.28 &    1a        \Tstrut\\
            &   7090            &      -             &      -        &         -         &       -           &     1b       \\
            &   6950            &    3.90            &      -        &         -         &       -           &      1c       \\
            &   7328            &      -             &      -        &         -         &     0.91          &      2        \\ 
            &   6800            &      -             &      -        &         -         &       -           &     3        \\
            &   7263            &      -             &      -        &         -         & 2.04\,$\pm$\,0.62 &      6        \\
            & 7072\,$\pm$\,210  &    3.36            &      -        &         -         &       -           &   8        \\
            & 6930\,$\pm$\,130  &      -             &      -        &         -         &       -           &      10a       \\
            & 7090\,$\pm$\,160  & 3.84$\,\pm$\,0.15  &      -        &         -         &       -           &     10b       \\
            & 6900\,$\pm$\,200  & 3.40$\,\pm$\,0.10  & 65\,$\pm$\,6  &  2.6\,$\pm$\,0.2  & 1.34\,$\pm$\,0.14 &    10c       \\
            &   6900            &    3.84            &      -        &         -         &       -           &     11       \\\hline 
HD\,118660  & 7550\,$\pm$\,150  & 4.00$\,\pm$\,0.10  &108\,$\pm$\,8  &  3.9\,$\pm$\,0.7  & 1.12\,$\pm$\,0.27 &   1a \Tstrut\\
            &   7610            &      -             &      -        &         -         &       -           &      1b       \\
            &   7460            &    3.95            &      -        &         -         &       -           &     1c       \\
            &   7638            &      -             &      -        &         -         &     0.96          &      2        \\
            &   7177            &      -             &      -        &         -         &       -           &     3        \\ 
            &   7500            &      -             &      -        &         -         & 1.03\,$\pm$\,0.05 &     6        \\ 
            & 7340\,$\pm$\,130  &      -             &      -        &         -         &       -           &     10a       \\
            & 7500\,$\pm$\,120  & 4.05$\,\pm$\,0.16  &      -        &         -         &       -           &      10b       \\
            & 7200\,$\pm$\,200  & 3.90$\,\pm$\,0.10  &100\,$\pm$\,10 &  2.4\,$\pm$\,0.2  & 0.98\,$\pm$\,0.14 &     10c       \\
            &   7772            &      -             &      -        &         -         &       -           &     12       \\\hline
\hline                                                                                   
\end{tabular}
}
\end{center}  
\begin{minipage}{\textwidth}%
Notes: First column is the star name, second to sixth columns are atmospheric parameters. The last column lists the source of measurements.
 (1a) High-resolution spectroscopic parameters from the present study;
 (1b) Photometric calibration from \citet{1985MNRAS.217..305M}; 
 (1c) Photometric calibration from \citet{1993A&A...268..653N};
 (2) \citet{2011MNRAS.414..792B}; (3) \citet{2012MNRAS.427..343M};
 (4) \citet{2002ChJAA...2..441X}; (5) \citet{2003MNRAS.344..431J};
 (6) \citet{joshi2009};
 (7) \citet{1999MNRAS.309..871M}; (8) \citet{casagrande2011};
 (9) \citet{smalley2011};         (10a) \citet{2014MNRAS.C} ($V-K_s$ colors);
 (10b) \citet{2014MNRAS.C} ($uvby\beta$ photometry);
 (10c) \citet{2014MNRAS.C} (spectroscopic);
 (11) \citet{2011MNRAS.411..435B}; (12) \citet{masana2006} 
\end{minipage}
\end{table*}		
		
		\subsection{Photometric Indicators}
		
		To determine \teff{} and \logg{} from the Str\"omgren photometric indices, we used the  calibrations of \citet{1985MNRAS.217..305M} and \citet{1993A&A...268..653N} implemented in \textsc{uvbybeta} program under IDL library \textsc{Astrolib} and FORTRAN program \textsc{uvbybeta\_new}, respectively. These calibrations are based on the Str\"omgren photometric indices $b-y$, $m_1$, $c_1$ and $\beta$.  Since the photometric sets of indices for HD\,98851 and HD\,102480 are not availble in the liturature, hence one of the missing indices $\beta$ was determined from their spectral class. The basic parameters obtained using photometric method are listed in Table~\ref{table:final_results} with reference 1b and 1c.

		\begin{figure}
			\begin{center}
				\includegraphics[width=0.45\textwidth]{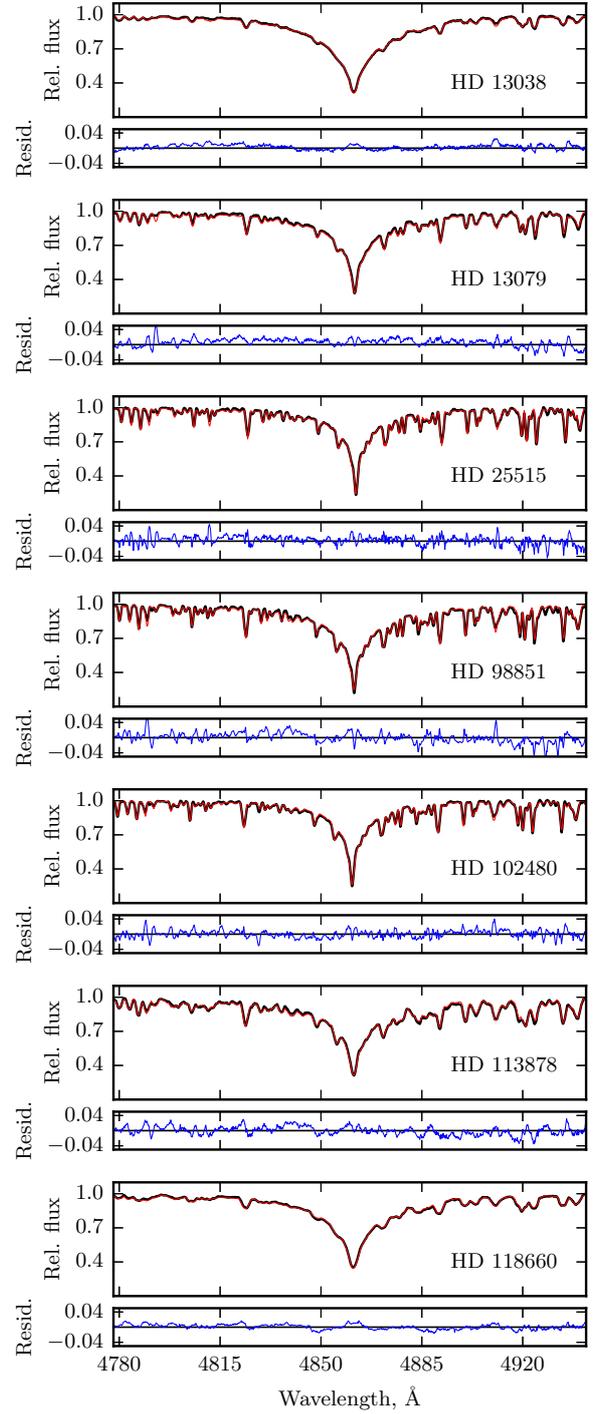}
				\caption{The upper panels show the best fit of an observed $H_\beta$ line profile (black) to synthetic (red). The lower panels show the residuals of the best fit.}
				\label{fig:hbetafit}
			\end{center}
		\end{figure}

		\subsection{Spectroscopic Indicators}
				
	 Apart from the photometric calibrations, spectroscopic method is widely used to measure the fundamental parameters independently. For this, we applied spectral fitting technique to the observed high-resolution spectra. To generate the synthetic spectra from the atmospheric models, initial guess of these parameters were obtained from the photometric calibration. Since for cooler stars (\teff\,$<$ 8000 K), the hydrogen lines are the best indicator of \teff{}, hence we used the $H_{\beta}$ line fitting technique implemented within IDL-based SME\footnote{\url{http://www.stsci.edu/\~valenti/sme.html}} (Spectroscopy Made Easy) software written by \citet{2016arXiv160606073P}. Varying the abundances of the selected elements and other parameters, this code does simultaneous fitting in various spectral regions. SME utilizes line lists from Vienna Atomic Line Database \citep[VALD3,][]{ryabchikova2015}. Each run of SME returns the values for unknown parameters and associated errors \citep{2016arXiv160606073P}. This routine automatically fits the observed spectra with synthetic ones assuming  hydrostatic, plane-parallel atmospheres and LTE approach. This package includes several sets of atmospheric models e.g. ATLAS9, ATLAS12, LL\_models, etc. In the present work, the stellar atmosphere model powered by ATLAS9 \citep{1979ApJS...40....1K} was used. On applying SME to the intensity spectra, we refined the basic parameters calculated through photometry. The Balmer line $H_{\beta}$ in MSS spectra was fitted for all the sample stars to determine \teff{} and Fig.~\ref{fig:hbetafit} demonstrates the goodness of fit. The surface gravity \logg{} was estimated from fitting of Mg I triplet line as shown in Fig.~\ref{figure:nes}. The atmospheric parameters obtained in this manner are listed in columns 4--8 of Table~\ref{table:final_results} with reference 1a.

	\section{Magnetic Field Measurements}
	\label{section:magnetic}
	
	Magnetic fields in early A-type stars are stable on a time scale of many decades and appear to be `frozen-in' to a rigidly rotating atmosphere. Though it is generally believed that spots in late type stars originate due to intense magnetic field, however, \citet{2013MNRAS.431.2240B} detected spots in some A-type stars where the magnetic field is either weak or absent. The strength of the magnetic field allows us to distinguish the two types of CP stars namely Ap (magnetic) and Am (non-magnetic) stars. Therefore, it is essential to check the presence of magnetic field in the sample stars before performing further spectroscopic analysis. Magnetic field can be detected through the Zeeman effect and due to this, the left and right circular polarization spectra are shifted with respect to each other which is proportional to the value of longitudinal magnetic field averaged over the stellar disk. The measured magnetic field  using this technique is known as the mean longitudinal or effective magnetic field \citep{1947PASP...59..112B,1958ApJ...128..228B} and can be measured directly from the polarized spectra.

	\begin{figure}
		\begin{center}
			\includegraphics[width=0.48\textwidth]{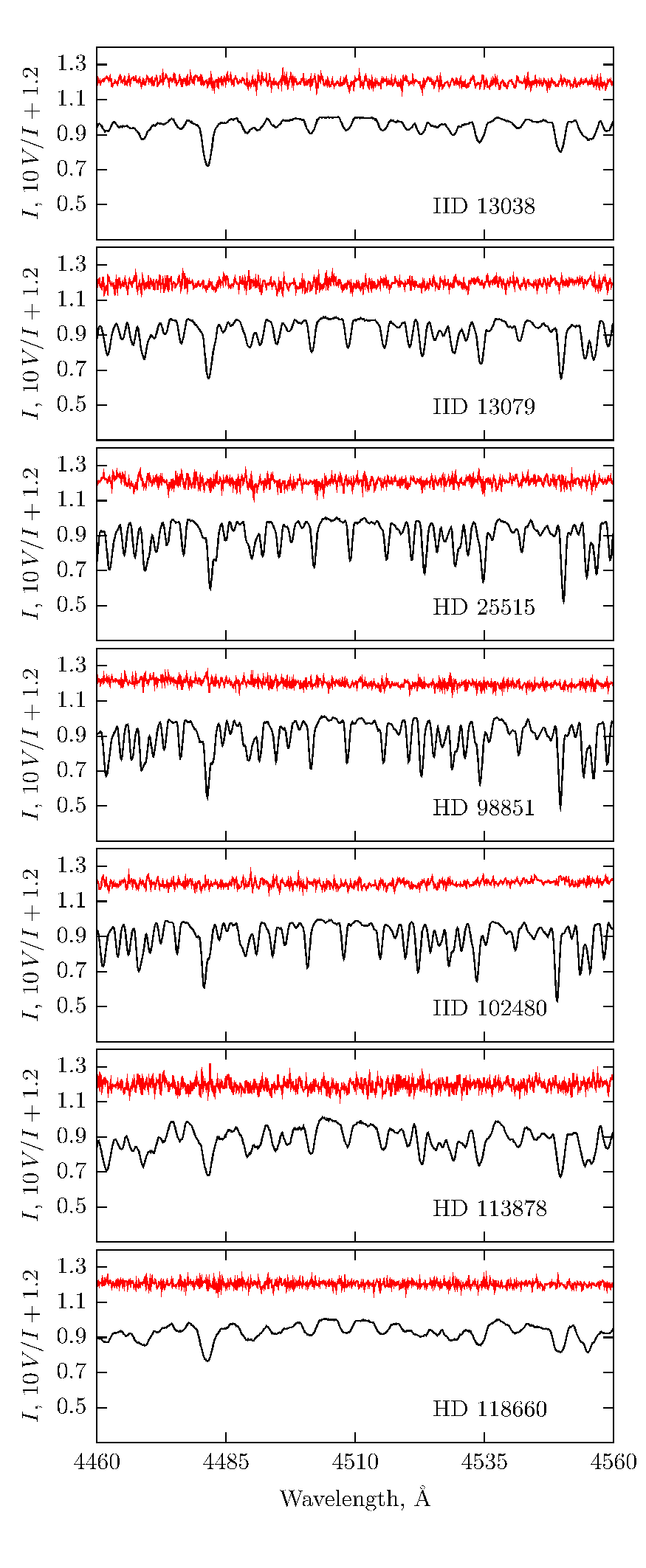}
			\caption{Intensity and circularly polarized spectra of sample stars as shown by black and red color, respectively.}
			\label{fig:polarim}
		\end{center}
	\end{figure}

	According to the basic principles of Zeeman effect, longitudinal magnetic field $B_\mathrm{z}$ expressed in  the unit of Gauss(G)  induces the splitting of a single line of initial wavelength $\lambda_\mathrm{0}$ and Land\'e factor $g$ into two circularly polarized $\sigma$-components (or groups of the components) separated by wavelength
	
	\begin{equation} \label{eq:zeeman}
	\Delta\lambda = 9.34\times 10^{-13}\,\times \lambda^2_\mathrm{0}\,\times g\,\times B_\mathrm{z}\,\, [\mathrm{\AA}]
	\end{equation}
	
The Zeeman shift $\Delta\lambda$ can be evaluated as the difference between the center of gravity (`COG') of two components or difference between two approximating functions to the lines under investigation. The fitting is usually done with a Gaussian function where the lines are non-distorted e.g. blending.
	
\begin{table}
		\caption{Measurements of longitudinal magnetic field strength of the sample stars. The error in measurements of the magnetic field is given by $\sigma$. The parameter $n$ is the total number of lines used for the analysis. For comparison, magnetic field measurements of known standard magnetic (marked with *) and non-magnetic stars (marked with **) are also given in the second half of this table.}
		\label{table:longfield}
		\begin{center}
			\scalebox{0.83}{
			\begin{tabular}{lcrrr}
				\hline
				Star       & JD (24500000+) & $B_{z}^\mathrm{COG}\pm\sigma$ (n) & $B_{z}^\mathrm{regres}\pm\sigma$ \\
				 &        (day)            &            (G)                    &        (G)    \\                
				\hline
				HD\,13038  & 6644.407        &  $1{\pm}215$ (56)      & $-50{\pm}55$   \\
				HD\,13079  & 6644.383        &  $-140{\pm}100$ (99)   & $-60{\pm}35$   \\
				HD\,25515  & 6644.490        &  $-30{\pm}70$  (135)   & $15{\pm}20$    \\
				HD\,98851  & 6383.463        &  $35{\pm}50$ (131)     & $40{\pm}15$    \\
				HD\,102480 & 6644.533        &  $-125{\pm}50$ (129)   & $-130{\pm}15$ \\
				HD\,113878 & 6381.497        &  $50{\pm}170$ (84)     & $-60{\pm}40$  \\
				          & 6383.486        & $30{\pm}100$ (73)      & $-25{\pm}30$   \\
				HD\,118660 & 6383.521        & $190{\pm}145$ (78)     & $180{\pm}45$   \\
				          & 6383.538        & $40{\pm}190$ (48)      & $0{\pm}75$     \\
				\hline
				$\alpha^2$ CVn* & 6381.512 & $-445{\pm}55$ (148)     & $-290{\pm}25$  \\
				                & 6383.503 & $-845{\pm}50$ (130)     & $-800{\pm}25$ \\
				53 Cam*         & 6644.506 &  $-3720{\pm}130$ (194)  & $-2970{\pm}40$ \\
				$o$ UMa**       & 6644.513 & $-8{\pm}50$ (293)       & $-10{\pm}10$    \\
				HD\,109317**     & 6381.530 & $25{\pm}50$ (292)       & $25{\pm}10$     \\
  				                & 6383.510 & $65{\pm}50$ (297)       & $65{\pm}10$     \\
				\hline
			\end{tabular}
		}
		\end{center}
	\end{table}
An alternative technique utilizes polarized spectra in the form of two Stokes parameters and description of this method can be found, for instance, in \citet{2002A&A...389..191B}. Circularly polarized signal or $V$ Stokes parameter in stellar spectra can be expressed as  

	\begin{equation} \label{eq:regres}
	\frac{V}{I} = -4.67\times 10^{-13}\,\times \lambda^{2}_{0}\,\times g\,\times \frac{1}{I}\,\times \frac{\mathrm{d}I}{\mathrm{d}\lambda}\,\times B_\mathrm{z}
	\end{equation}

		\begin{figure*}
			\centering
			\includegraphics[width=0.8\textwidth]{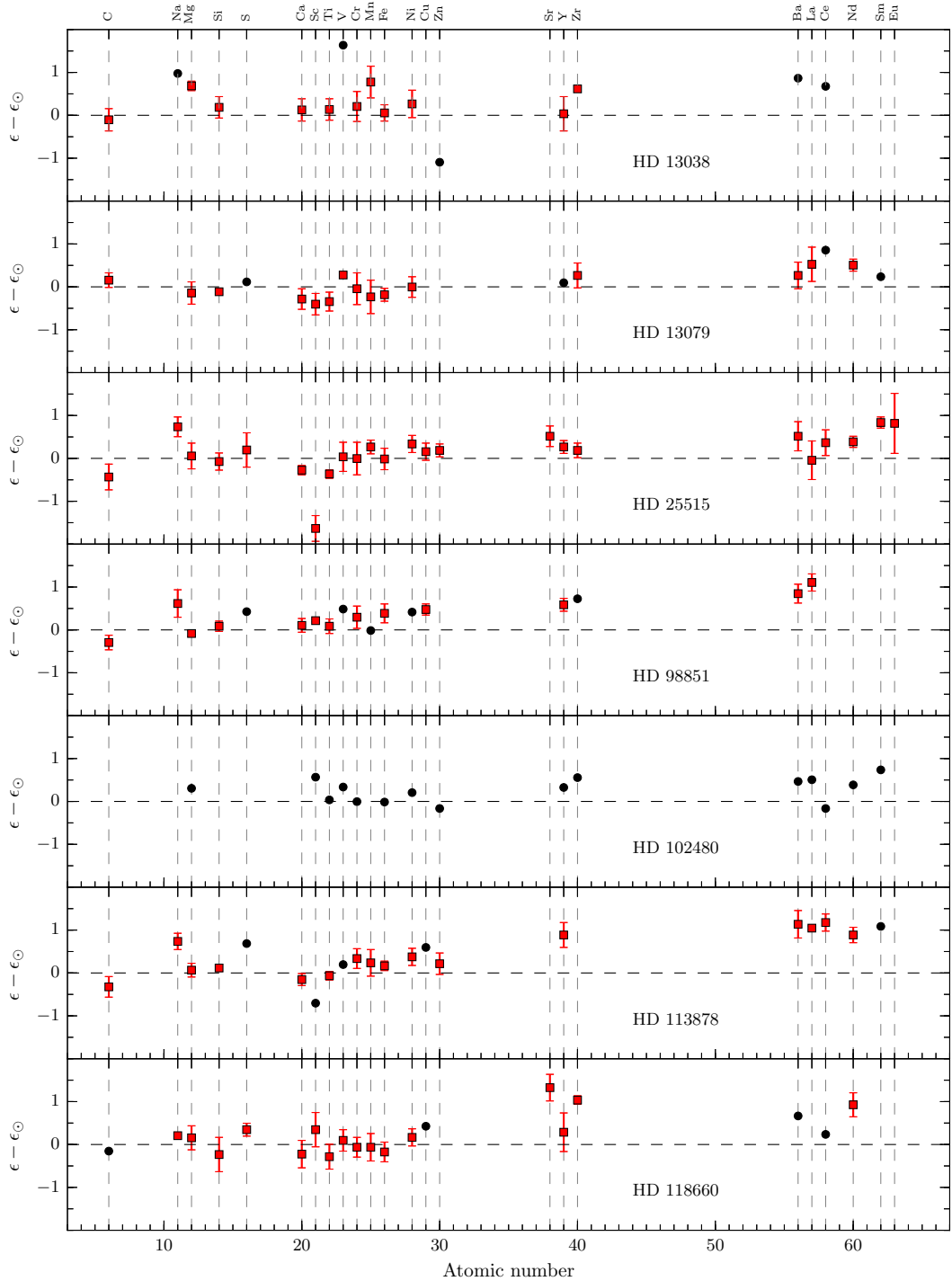}
			\caption{The elemental abundances of sample stars relative to the Sun.}
			\label{fig:abundance-by-atom.eps}
		\end{figure*}

The $V$ Stokes parameter is defined as difference between right and left handed circularly polarized spectra and intensity, $I$, is the half of the sum of intensity of right and left handed circularly polarized light. The strength of mean longitudinal field $B_\mathrm{z}$ in Eq. \ref{eq:regres} is the coefficient of linear regression. Hereafter, we denote, the result obtained though this method with an upper index `regres'. Eq. \ref{eq:regres} becomes more important when magnetic field is weak and Zeeman broadening is negligible relative to other mechanisms of broadening. With a prior knowledge that all the sample stars are classified as Am stars, hence principally either they do not posses any magnetic field, or if the magnetic field is presented the strength should be weak. Therefore, in addition to the common method `COG', we apply the `regres' method to our polarized spectra for the detection of weak magnetic fields. For both the methods the mean adopted value of effective Land\'e factor is 1.23 \citep{1984AISAO..18...37R} which is very close to 1.21 as given by \citet{2002A&A...389..191B}. The intensity and circularly polarized spectra of the sample stars are depicted in Fig.~\ref{fig:polarim} and the strength of measured magnetic field  is summarized in Table~\ref{table:longfield}.
	
\begin{table*}
			\caption{The photospheric chemical abundances ($\varepsilon$=$log(N_{cl}/N_{tot})$) of the sample stars. Colon (:) sign denotes doubtful measurements. The solar abundance $\varepsilon_o$ \citep{2007SSRv..130..105G} of the corresponding element is also listed in the last column.}
			\label{table:table4}
			\tiny
			\begin{center}
				{\small
					\begin{tabular}{lrrrrrrrr}
						\hline
						Atom           & HD\,13038      & HD\,13079      & HD\,25515                 & HD\,98851        & HD\,102480 & HD\,113878     & HD\,118660     &  Sun    \\
						\hline
						C               & $-3.75\pm0.26$ & $-3.49\pm0.17$ & $-4.08\pm0.30$ & $-3.94\pm0.17$   &   $-$      & $-3.97\pm0.24$ & $-3.80$ (:)    & $-3.65$ \\
						Na              & $-4.89$ (:)    &     $-$        & $-5.13\pm0.23$ & $-5.25\pm0.32$   &   $-$      & $-5.13\pm0.19$ & $-5.66\pm0.08$ & $-5.87$ \\
						Mg              & $-3.82\pm0.11$ & $-4.65\pm0.26$ & $-4.45\pm0.30$ & $-4.59\pm0.06$   &  $-4.20$   & $-4.44\pm0.16$ & $-4.35\pm0.28$ & $-4.51$ \\
						Si              & $-4.34\pm0.25$ & $-4.64\pm0.05$ & $-4.60\pm0.20$ & $-4.44\pm0.12$   &   $-$      & $-4.41\pm0.08$ & $-4.76\pm0.40$ & $-4.53$ \\
						S               & $-$            & $-4.78$ (:)    & $-4.70\pm0.40$ & $-4.47(:)$       &   $-$      & $-4.21$ (:)    & $-4.55\pm0.15$ & $-4.90$ \\
						Ca              & $-5.60\pm0.26$ & $-6.01\pm0.24$ & $-6.00\pm0.11$ & $-5.62\pm0.16$   &   $-$      & $-5.88\pm0.14$ & $-5.95\pm0.32$ & $-5.73$ \\
						Sc              & $-$            & $-9.27\pm0.25$ & $-10.50\pm0.30$ & $-8.65\pm0.03$  &  $-8.30$   & $-9.57$ (:)    & $-8.52\pm0.40$ & $-8.87$ \\
						Ti              & $-7.00\pm0.25$ & $-7.48\pm0.22$ & $-7.50\pm0.10$ & $-7.05\pm0.17$   &  $-7.10$   & $-7.20\pm0.10$ & $-7.42\pm0.29$ & $-7.14$ \\
						V               & $-6.40$ (:)    & $-7.76\pm0.05$ & $-8.00\pm0.34$ & $-7.55$ (:)      &  $-7.70$   & $-7.84$ (:)    & $-7.94\pm0.25$ & $-8.04$ \\
						Cr              & $-6.19\pm0.35$ & $-6.44\pm0.37$ & $-6.40\pm0.38$ & $-6.10\pm0.26$   &  $-6.40$   & $-6.06\pm0.23$ & $-6.46\pm0.23$ & $-6.40$ \\
						Mn              & $-5.87\pm0.37$ & $-6.88\pm0.39$ & $-6.38\pm0.16$ & $-6.66$ (:)      &   $-$      &  $-$           & $-6.71\pm0.32$ & $-6.65$ \\
						Fe              & $-4.53\pm0.19$ & $-4.77\pm0.15$ & $-4.60\pm0.25$ & $-4.20\pm0.22$   &  $-4.60$   & $-4.42\pm0.11$ & $-4.76\pm0.23$ & $-4.59$ \\
						Ni              & $-5.54\pm0.32$ & $-5.81\pm0.24$ & $-5.47\pm0.20$ & $-5.39\pm0.25$   &  $-5.60$   & $-5.43\pm0.20$ & $-5.64\pm0.20$ & $-5.81$ \\
						Cu              & $-$            & $-$            & $-7.67\pm0.20$ & $-7.35\pm0.13$   &   $-$      & $-7.23$ (:)    & $-7.40$ (:)    & $-7.83$ \\
						Zn              & $-8.53$ (:)    & $-$            & $-7.25\pm0.15$ & $-$              &  $-7.60$   & $-$            & $-$            & $-7.44$ \\
						Sr              & $-$            & $-$            & $-8.60\pm0.24$ & $-$              &   $-$      & $-$            & $-7.79\pm0.31$ & $-9.12$ \\
						Y               & $-9.79\pm0.40$ & $-9.73$ (:)    & $-9.56\pm0.15$ & $-9.24\pm0.15$   &  $-9.50$   & $-8.94\pm0.29$ & $-9.54\pm0.45$ & $-9.83$ \\
						Zr              & $-8.84\pm0.03$ & $-9.19\pm0.29$ & $-9.27\pm0.17$ & $-8.73$ (:)      &  $-8.90$   &    $-$         & $-8.42\pm0.10$ & $-9.46$ \\
						Ba              & $-9.00$ (:)    & $-9.60\pm0.31$ & $-9.35\pm0.34$ & $-9.02\pm0.22$   &  $-9.40$   & $-8.57$ (:)    & $-9.20$ (:)    & $-9.87$ \\
						La              & $-$            & $-10.38\pm0.40$& $-10.95\pm0.45$ & $-9.80\pm0.20$   &  $-10.40$  & $-9.86\pm0.05$ & $-$            & $-10.91$ \\
						Ce              & $-9.66$ (:)     & $-9.48$ (:)    & $-9.97\pm0.30$ & $-$              &  $-10.50$  & $-9.16\pm0.20$ & $-10.10$ (:)   & $-10.34$ \\
						Nd              & $-$            & $-10.08\pm0.14$& $-10.20\pm0.13$ & $-$              &  $-10.20$  & $-9.70\pm0.18$ & $-9.66\pm0.28$ & $-10.59$ \\
						Sm              & $-$            & $-10.80$ (:)   & $-10.70\pm0.70$ & $-$              &  $-10.30$  & $-$            & $-$            & $-11.04$ \\
						\hline
					\end{tabular}
				}
			\end{center}
		\end{table*}

Apart from the target stars, on each night, we also observed a set of magnetic and non-magnetic standard stars. For our convenience, we divided them into two groups. The first one includes the magnetic stars of known period of magnetic variability. $\alpha^2$ CVn and 53 Cam are two well known examples of magnetic variables of periods 5.46939 and 8.02681 days, respectively \citep{2002APN....38.....W}. The second group consists of non-magnetic stars used to control the value of instrumental polarization. In our observations, $o$ UMa and HD\,109317 were selected as control stars. Unlike the magnetic standards, these  stars have no signatures of magnetic field within the typical errors of $B_z$ measurements. Hence, any possible  detection of $B_{z}$ in non-magnetic stars is attributed to the instrumental end. The main source of errors in $B_{z}$ measurements associates with spectra of poor quality (low S/N), instrumental polarization from telescope, associated devices and the accuracy of guiding that mostly depends upon the weather conditions.

On carefully analyzing the results summarized in Table~\ref{table:longfield}, we conclude that none of the sample star has significant magnetic field strength. The point worth to be noted here is the better accuracy in the determination of effective magnetic field via `regres' method.  Depending on the spectral resolution, SNR and spectral coverage, this method is useful in measuring the weak magnetic field. However, this method is more sensitive to the accuracy of continuum normalization. We applied this method to the spectra obtained from MSS and compared with the classic `COG' method and we found that the difference between $B_{z}^\mathrm{COG}$ and $B_{z}^\mathrm{regres}$ is negligible in case of narrow spectral profiles.

\section{Abundance Analysis} 
\label{section:abundance}

After determining the basic parameters, the chemical abundance analysis was carried out using spectrum synthesis technique by fitting of synthetic spectra to the observed one. To derive the abundance of chemical elements we applied SME~ to NES \'{e}chelle spectra. The output of each run of SME returns the values of unknown parameters such as surface chemical abundances, $v\sin i$ etc. and errors associated with them. We have listed the photospheric elemental distribution $\log(\frac{N_\mathrm{el}}{N_\mathrm{tot}})$ of studied star in Table \ref{table:table4} and graphically represented in Fig. \ref{fig:abundance-by-atom.eps}.

\begin{figure}
\centering
\includegraphics[width=\columnwidth]{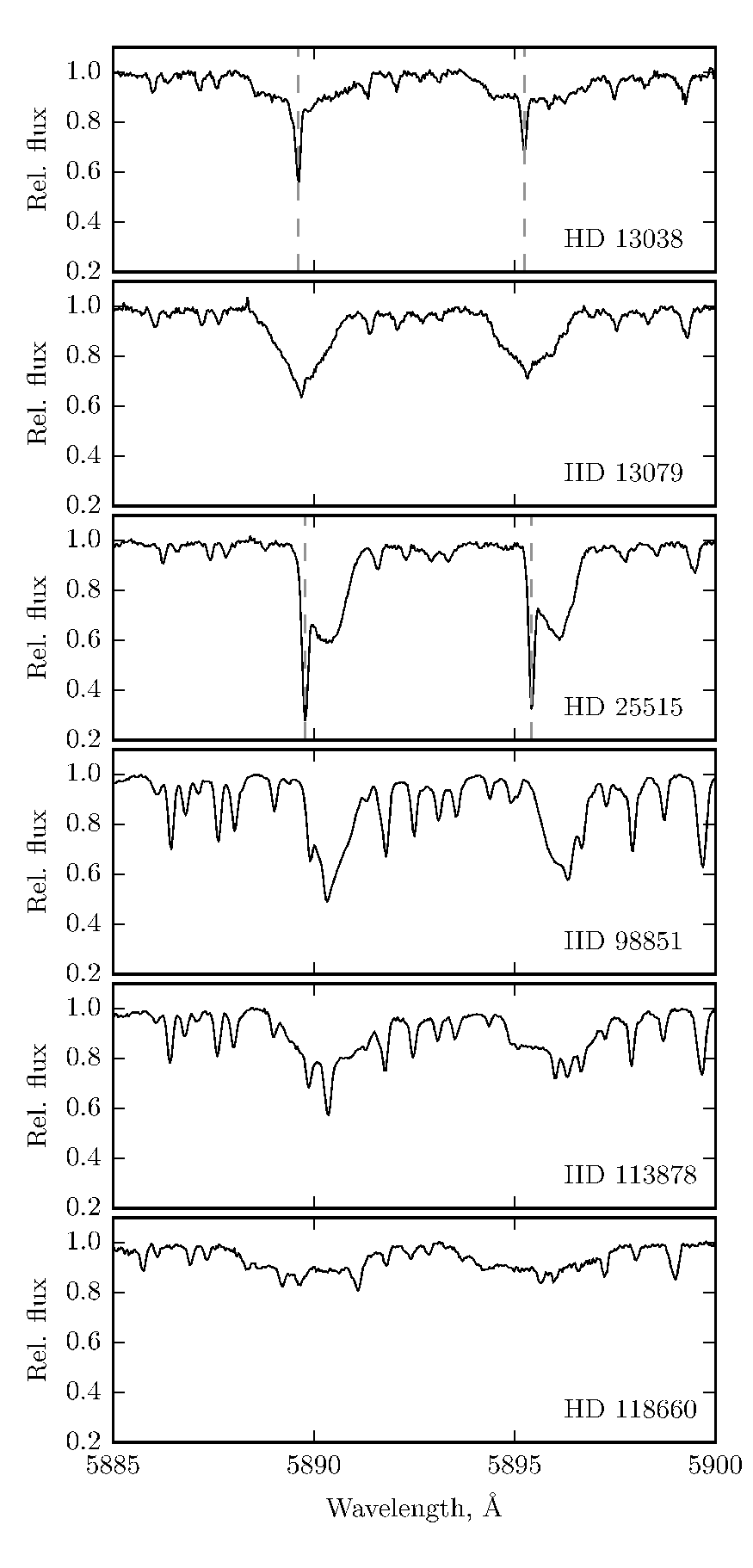}
\caption{Spectra of program stars in the region of sodium doublet Na D1 and Na D2. The marked lines are correspond to the interstellar components having significant extinction ($>$0.1 mag). Spectrum of those stars where the sodium doublets are blended have not been marked.}
\label{figure:sodium}
\end{figure}

Features of the light, iron peak and heavy elements are of particular interest in CP stars because the non-solar abundance of these elements distinguishes them with the normal star. As mentioned in the introduction, Am stars exhibit overabundance of the iron peak and heavy elements such as Zn, Sr, Zr, Ba and underabundace of the elements such as Ca, Sc \citep{2009ssc..book.....G}. Therefore, our first aim was to select the least blended lines of these elements and analyze them. It can be seen from Table~\ref{table:table4} that  almost in each star Ca, Sc, and some light elements show mild deficiency or solar value while iron and heavy elements are excess in comparison to the solar abundances \citep{2007SSRv..130..105G}. We have also found that in some of sample stars these elements have abundances similar to Sun within 0.4 dex. Based on their derived chemical abundances we found that four of sample stars (HD\,13079, HD\,25515, HD\,113878 and HD\,118660) belong to the category of mild Am star and rest of them (HD\,13038, HD\,98851 and HD\,102480) to the class of normal Am star. Due to slow rotation, the spectra of Am stars are usually characterized by the narrower profiles in comparison to their normal counterpart. On investigating their rotational velocity it is found that all the sample stars have rotational velocity less than 90 km\,s$^{-1}$ except HD\,118660 with $v \sin i$ of 108 km\,s$^{-1}$ and this is reflected in the form of broadening of the spectrum of this star (see, Fig \ref{figure:nes).}

\section{Evolutionary Status}\label{sec:hrd}
\label{section:evolution}

In order to know the evolutionary status of a star, its position in the H-R diagram should be known precisely. For this, one should have the accurate values of effective temperature (\teff{}) and luminosity ($\log (\rm{L/L_{\odot}})$) because a typical error of 150\,K in the determination of \teff{} leads to an error of about 0.20 mag in the bolometric magnitude \citep{2011MNRAS.411.1167C, 2015MNRAS.454L..86N}. The basic parameters such as \teff{} were calculated using both the photometric and spectroscopic methods but for constructing the H-R diagram, we have adopted \teff{} derived through spectroscopy (Table~\ref{table:final_results}). 
	
The luminosity $\log (\rm{L/L_{\odot}})$ of a star can be calculated using a general relation
	
\begin{equation}
\log \left (\frac{L_\star}{L_\odot} \right ) = - \frac{M_v + BC - M_{bol,\odot}} {2.5},
\label{eq:eq1}
\end{equation}
	
Absolute V magnitude can be computed from the relation, $M_v = m_v + 5 + 5\,\log\, \pi - A_v$ using the data given in Table~\ref{table:samples}. Bolometric corrections ($BC$) for all the sample stars are computed by `spline' interpolation using the grids of \citet{1996ApJ...469..355F} over $\log$ \teff{} values and $M_{bol,\odot}\,=\,4.74$.  

Visual absorption, $A_V=3.1\,\times\,E(B-V)$ \citep{1979ARA&A..17...73S} is an important parameter which is insignificant for nearby stars and those are located at high galactic latitude. Since all the target stars are close by and most of them lie on high galactic latitude (see Table \ref{tab:extinction}), hence we could have ignored their extinction values.
Nevertheless, we used four different approaches to derive $A_v$ values those are summarized in Table~\ref{tab:extinction}.
\citet{lucke78} and \citet{schlegel98} provide a three-dimensional maps of galactic dust while \citet{green2015} provide interstellar reddening based on photometric data taken from 2MASS \citep{2006AJ....131.1163S} and Pan-STARRS 1 \citep{2010SPIE.7733E..0EK} database. Application of the calibration of \citet{1997A&A...318..269M} on the high-resolution spectra was also used to evaluate the interstellar extinction. Fig.~\ref{figure:sodium} shows the spectra of sample stars covering  Na~D region those was used to measure the extinction parameters. 

\begin{table*}
\caption{The visual extinction of the sample stars estimated through interstellar extinction maps and determined using high-resolution spectroscopy.}
\label{tab:extinction}
\begin{center}
\begin{tabular}{lrrccccc}
\hline
Star    &  $l$ & $b$ & \multicolumn{5}{c}{Extinction values ($A_v$)}   \\
     & & &  \citet{lucke78} & \citet{schlegel98} & \citet{green2015} &  Spectroscopy & Average  \\
\hline
HD\,13038   &   133.16  & $-3.35$  &    0.37     &  -     &     -      & 0.31  & 0.34$\pm$0.03 \\
HD\,13079   &   138.83  & $-20.87$ &      -      &  -     &   0.03     & 0.09  & 0.06$\pm$0.03 \\
HD\,25515   &   151.24  & $-1.12$  &    0.34     &  -     &   0.34     & 0.31  & 0.33$\pm$0.01 \\
HD\,98851   &   193.89  & $+70.20$ &      -      &  0.06  &   0.03     &  -    & 0.04$\pm$0.02 \\
HD\,102480  &   143.16  & $+61.55$ &      -      &  0.09  &   0.02     &  -    & 0.06$\pm$0.04 \\
HD\,113878  &   116.15  & $+68.91$ &    0.25     &  0.03  &   0.04     &  -    & 0.11$\pm$0.10 \\
HD\,118660  &   345.58  & $+73.18$ &     -       &  0.09  &   0.01     &  -    & 0.05$\pm$0.04 \\
\hline
\end{tabular}
\end{center}
\end{table*}

The spectrum of HD\,13038 shows strong interstellar component of Na doublet. The measured equivalent width (EW) of Na D1 line in the spectrum of this star gives $A_v$\,$\sim$\,0.31 mag which is close to 0.37 mag that derived through the maps of \citet{lucke78}. For HD\,13079, due to significant blending effects the measurement of the EW was difficult. However for this star, we placed an upper value of $A_v\approx$\,0.09 mag. According to the maps of \citet{lucke78}, HD\,13079 is located in a region free from extinction, hence we can ignore the extinction for this star. HD\,25515 shows the strongest interstellar lines in its spectrum. For this star, the extinction maps of \citet{lucke78} and \citet{green2015} revealed the value of $A_v$\,=\,0.34 mag which is close to 0.31 mag, a value obtained through the measurement of EW of Na D1 line.
For the stars HD\,98851, HD\,102480 and HD\,118660, extinction is negligible as they are located high above the galactic plane and the same is also true from the extinction maps. {\'{E}chelle} spectra of  HD\,98851 and HD\,118660 shows some signatures of interstellar Na \textsc{I} lines but they are too weak and strongly blended, hence making difficult to measure their EWs. For HD\,102480, we do not have the spectra in the region of Na doublet, hence could not report the spectroscopic extinction. For star HD\,113878, the maps of \citet{lucke78} gives $A_v$\,=\,0.25 mag. Again, for this star the measurement of EWs of Na D lines became difficult due to strong blending effects. 
Table \ref{tab:extinction} lists the values of $A_v$ for sample stars  obtained from three extinction maps and high-resolution spectroscopy. The last column of this table gives average values of $A_v$ along with the associated errors and these are the values adopted for the computation of \luminosity{} values listed in the sixth column of Table \ref{table:final_results}.

\begin{figure}
\begin{center}
\includegraphics[width=0.5\textwidth,height=0.5\textwidth]{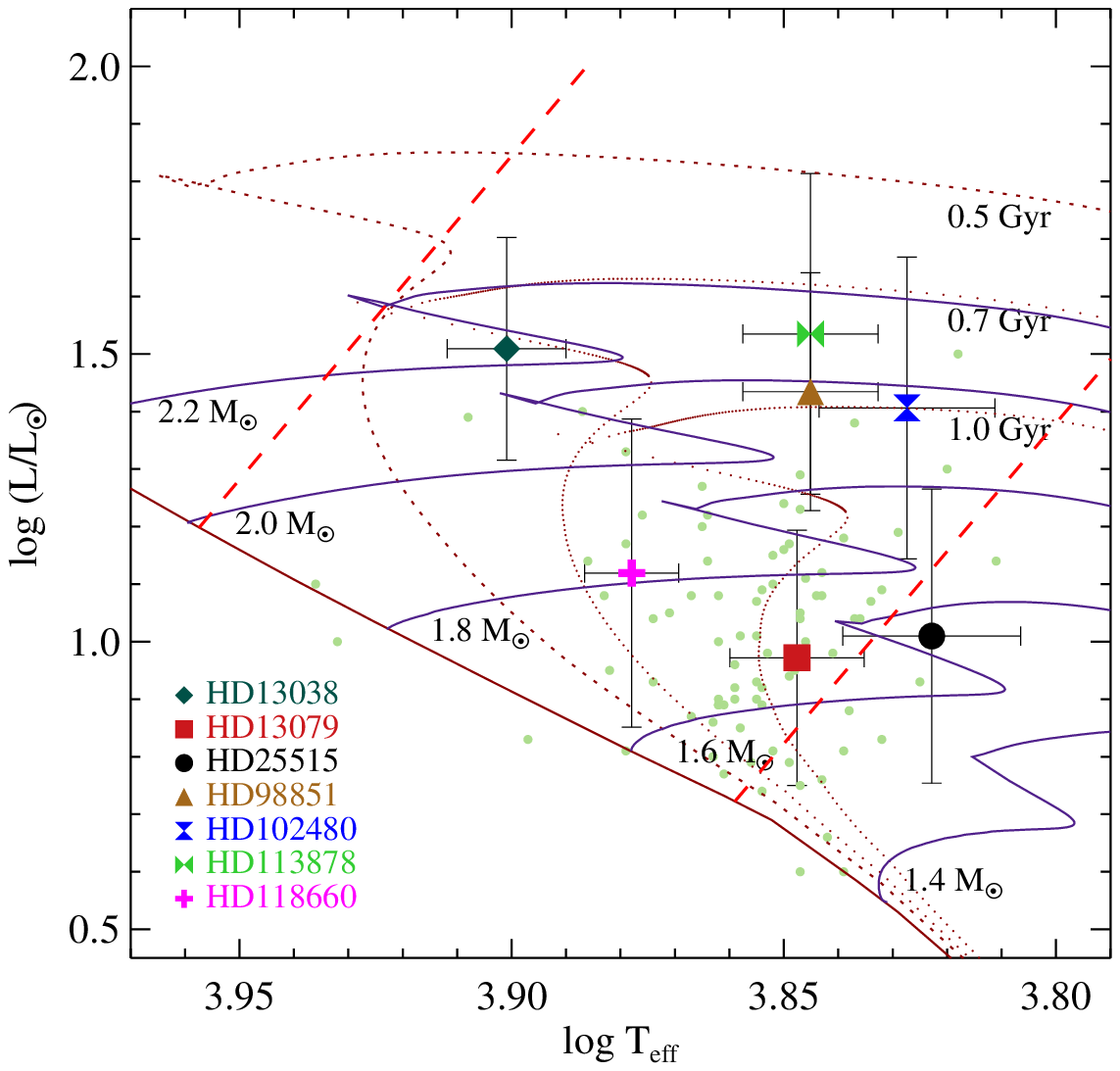}
\caption{Position of sample stars in H-R Diagram. Solar scaled  isochrones (dotted red line) and evolutionary tracks (solid purple line) are taken from BaSTI database. A sample of 87 Am stars with known $\log$ (\teff{}) and $\log (\rm{L/L_{\odot}})$ taken from \citet{smalley2011} are shown with light green color and small filled circles.  Our sample stars are displayed by different symbols as indicated in legend. The left and right red dashed lines mark the boundary of $\delta$ Sct instability strip \citep{breger1998}.}
\label{fig:hrd}
\end{center}
\end{figure}

For star HD\,13038, parallax is not available, hence the luminosity could not be derived through Eq. \ref{eq:eq1}. However for this star we computed the luminosity parameter using luminosity-mass empirical relationship expressed in terms of \teff{} and \logg{} for near solar metallicity Z=0.014 \citep{catanzaro2015},
	
\begin{eqnarray}
	\label{eq:eq2}
	\log \left (\frac{L_\star}{L_\odot} \right ) = -(14.909\,\pm\,0.006)+(5.406\,\pm\,0.002) \log T_{\rm eff}\\
	-(1.229\,\pm\,0.001) \log g~~({\rm rms=0.042~dex.}) \nonumber
	\end{eqnarray}
	
 Using this relation, we also derived \luminosity{} for other sample stars to to perform comparison between two techniques. Fig. \ref{fig:hrd} shows the locations of target stars in the H-R diagram. The  evolutionary tracks of masses 1.4, 1.6, 1.8, 2.0, 2.2 $M_{\odot}$ along with the isochrones of ages 0.5, 0.7 and 1.0 Gyrs taken from BaSTI database\footnote{\url{http://albione.oa-teramo.inaf.it/query_form.php}} for solar metalicity Z\,=\,0.0198 and Y\,=\,0.273 are also over-plotted. For the comparison, we have also plotted 87 Am stars for those $\log$ (\teff){} and $\log (\rm{L/L_{\odot}})$ values are available \citep{smalley2011}. The $\delta$ Sct instability strip taken from \citet{breger1998} is also shown by two slant lines (in red color). From this figure, one can conclude that all the sample stars have masses between 1.5 -- 2.5 $M_{\odot}$ and are evolved from Zero-Age-Main-Sequence (ZAMS) towards Terminal-Age-Main-Sequence (TAMS) or slightly beyond it.

\section{Comments on Individual Stars}\label{sec:individual_stars}
\label{section:star}
	
In the past, the target stars were studied using time-resolved photometric and low-resolution spectroscopic techniques while few of them were also investigated in high-resolution spectroscopic mode. The aim of these efforts  were to establish the pulsational phenomena for their asteroseismic investigation and ascertain various physical phenomena in the presence of their chemical peculiarities \citep{2004A&A...423..705R}. In following subsections, we provide brief information about the individual star along with their basic physical properties.
	
	\subsection{HD\,13038}
	
	This star is classified as A3 type in the HD catalog. Using the $uvby\beta$ photometric calibrations, \citet{2011MNRAS.414..792B} determined \teff{}\,=\,8147\,K. \citet{2002ChJAA...2..441X} found
\teff{} and $\log (\rm{L/L_{\odot}})$\ as 8210\,K and \,0.79, respectively. The $\delta$ Sct type pulsation with two close periods 28-min and 34-min were discovered by \citet{1999IBVS.4677....1M}. Previously, the peculiarity class of this star was not known. Our high-resolution spectroscopic analysis confirmed that most of the elements Ca, Fe and Cr have abundances similar to the Sun. The absence of magnetic field ascertains the peculiar type of this star as Am class.
	
\subsection{HD\,13079}

		HD\,13079 was discovered as a single mode $\delta$ Sct type pulsating variable with period 73-min~\citep{1999MNRAS.309..871M}. This star was further monitored under the SuperWASP survey\footnote{\url{http://www.superwasp.org/}} for which \citet{smalley2011} detected five pulsational frequencies. On fitting the H$\alpha$ line profile to the medium resolution (R$\sim$18000) spectrum, \citet{1999MNRAS.309..871M} calculated \teff{}\,=\,7200\,K and \logg{}\,=\,4.0. The value of photometric  temperature and luminosity determined by \citet{2002ChJAA...2..441X} are 7370\,K and 0.68, respectively. Using the Str\"{o}mgren photometric indices, \citet{smalley2011} determined $\log$ \teff\,=\,3.85 and $\log (\rm{L/L_{\odot}})$\,=\,0.74. Through the photometric calibration, \citet{2011MNRAS.414..792B} obtained  \teff{}\, as 7482\,K. The fitting of spectral energy distribution (SED) resulted  \teff{} of this star as 7270\,K \citep{2012MNRAS.427..343M}. Based on the analysis of low-resolution spectrum, \citet{1999MNRAS.309..871M} found that the abundance of iron peak elements are of solar value and Ca and Sc are underabundant. Our high resolution spectroscopic analysis revealed that Ca and Sc are marginally underabundant while heavy elements have abundances close to the solar values except La and Ce those are slightly overabundant. The spectropoloarimetric analysis resulted that the strength of magnetic field in this star is negligible that further supports its mild Am peculiarity.
	
\subsection{HD\,25515}

Using classical spectroscopy, \citet{1965PASP...77..184C} reported HD\,25515 as an Am star. \citet{2003AJ....125.2531R} classified this star as spectral type F3III. Pulsation in this star were discovered by \citet{2005BASI...33..371C} and later confirmed by \citet{joshi2009} who reported the presence of $\delta$ Sct type pulsational variability of period  2.78 hr. The SED fitting on the photometric data resulted \teff{}\,=\,6461\,K ~\citep{2012MNRAS.427..343M}. The photometric calibration of \citet{1985MNRAS.217..305M} lead values of  \teff{}, and \logg{} as \,6970\,K and \,3.46, respectively. Based on the high-resolution spectroscopy, we derived the values of  \teff{} and \logg{} as \,6650\,K and \,3.80, respectively. The abundance analysis from our high-resolution spectroscopic data shows that Sc is underabundant while heavy elements are overabundant. The abundance patterns and absence of magnetic field confirm the peculiarity of HD\,25515 as mild Am class.
	
	\subsection{HD\,98851}	
	
		The spectral type of HD\,98851 is F2  \citep{1980A&AS...39..205O}. \citet{1984ApJ...285..247A} classified this star as a marginal Am. The luminosity class of this star is designated as F3III by \citet{1999A&AS..137..451G}. Based on the low-resolution spectroscopy, \cite{2003MNRAS.344..431J} determined \teff{} and \logg{}  7000$\pm$250\,K and 3.5$\pm$0.5, respectively. Using photometric calibration, \citet{2011MNRAS.414..792B} determined \teff{}\,=\,6999\,K and \logg{}\,=\,3.50. On application of SED fitting, \citet{2012MNRAS.427..343M} derived effective temperature as 6800\,K. \citet{2003MNRAS.344..431J} found HD\,98851 pulsates with two periods of ratio 2:1 with unusual feature of alternative high- and low-maxima and sub-harmonic signature. Our high-resolution spectroscopic analysis indicate that most of the elements are close to the solar value and absence of magnetic field categorize this star as an Am class.

	\subsection{HD\,102480}
	
		This star has spectral type F5 \citep{1980A&AS...39..205O} and classified as Am star by \citet{1999A&AS..137..451G}. Based on the analysis of low-resolution spectroscopic data, \citet{2003MNRAS.344..431J} determined \teff{}\,=\,6750$\pm$250\,K, \logg{}\,=\,3.0$\pm$0.5 and $\log (\rm{L/L_{\odot}})$\,=\,1.4$\pm$0.2. \citet{2011MNRAS.414..792B} calculated \teff{} and \logg{} as 6966\,K and 3.11, respectively. The spectroscopic parameters \teff{}, \logg{} and \feh{} reported by \citet{casagrande2011} are 7000\,K, 3.80 and 0.17, respectively. The SED fitting technique resulted  \teff{}\,=\,6967\,K \citep{2012MNRAS.427..343M}. The unusual pulsations with alternative high- and low-maxima  with period ratio of 1:2 was detected by \cite{2003MNRAS.344..431J}. For this star, we could obtain only one spectrum from MSS spectropolarimeter and  analysis of this spectrum asserts its non-magnetic nature. The near solar values of the abundances obtained through the  medium resolution spectroscopy indicate that this star belongs to the group of Am star. The results of abundance analysis presented here are preliminary and in near future the high-resolution spectra are proposed to confirm the peculiar nature of this star. 
	
	\begin{figure*}
\centering
\includegraphics[width=0.8\textwidth]{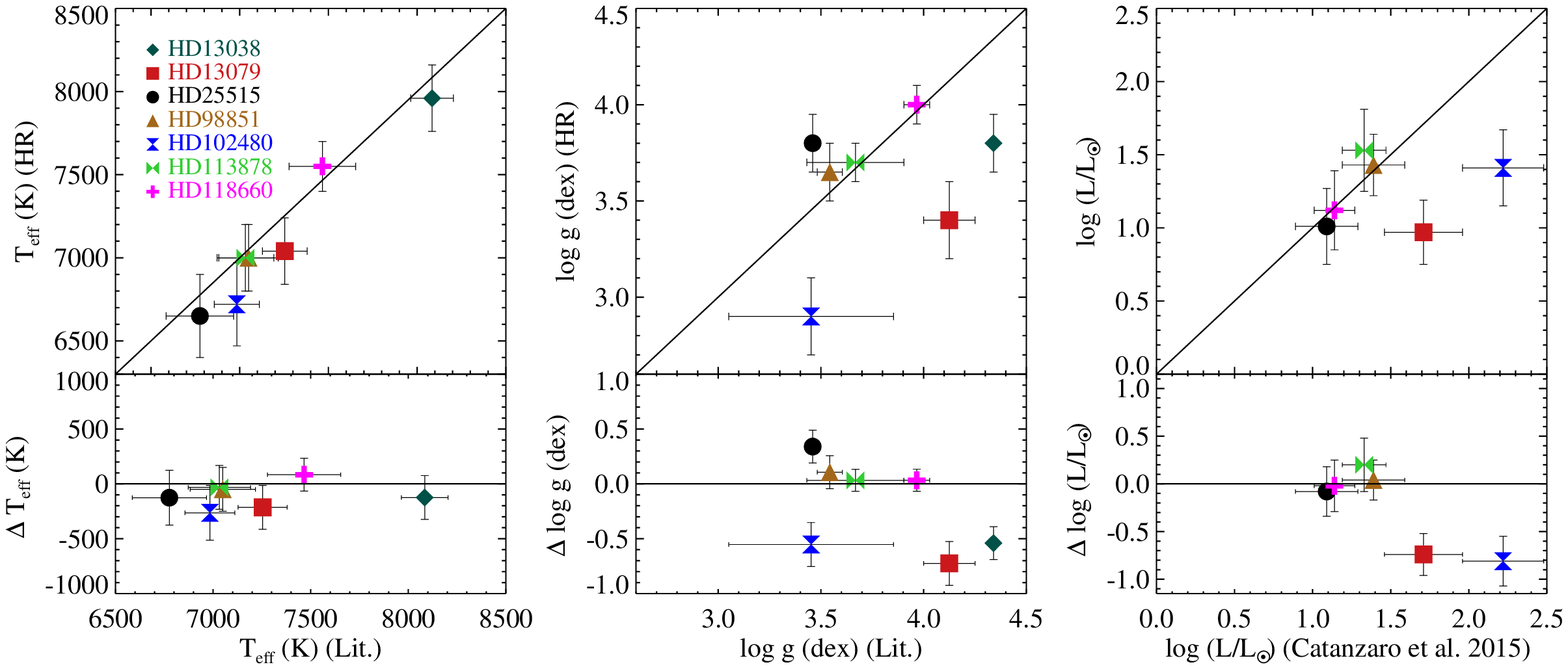}
\caption{Comparison of derived physical parameters \teff{} (left), \logg{} (center) and \luminosity{} (right) to average literature values listed in Table~\ref{table:final_results}. Abscissa display the mean values (in case of \luminosity, the value is derived using Eq.~\ref{eq:eq2}). Ordinates of upper panels show the derived values from high-resolution spectroscopy and lower panels their errors (Current work\,-\,Literature). The 1-$\sigma$ level is considered as the error in the compiled values. In the literature, for stars HD\,13038 and HD\,25515 we did not find errors in \logg\, hence the associated errors are not plotted in the center panels.}
\label{fig:scatter_comp}
\end{figure*}
	
	\subsection{HD\,113878}
	
		HD\,113878 is of A5 spectral type with metallic lines as discovered by \citet{1980A&AS...39..205O}  and classified as Am star by \citet{1999A&AS..137..451G}. From the photometric calibrations of \cite{1985MNRAS.217..305M} the calculated  value of \teff{}\, for this star is \,7263\,K \citep{joshi2009}.  \citet{2011MNRAS.414..792B} and \citet{2012MNRAS.427..343M} derive the \teff{}\, value as 7328\,K and 6800\,K, respectively. Using the Bayesian analysis technique, \citet{2011MNRAS.411..435B} calculated \teff{}\, and \logg{} as 6900\,K and 3.84, respectively. \citet{casagrande2011} determined the value \teff{}, \logg{} and \feh{} as 7072\,K, 3.36\, and 0.74\,, respectively. Using color index and photometric calibration, \citet{2014MNRAS.C} obtained \teff{} as 6930\,K and 7090\,K, respectively. This star was reported as a $\delta$ Sct type pulsating variable by \citet{2006A&A...455..303J}. Analysis of our high-resolution spectroscopic data revealed that HD\,113878 has deficiency of Ca, Sc, and Ti while abundance of heavy elements are in excess. The absence of magnetic field and pattern of elemental abundances support the peculiarity of HD\,113878 as mild Am class.
	
	\subsection{HD\,118660}
	
HD\,118660 is classified as A9 III-IV by \citet{1967PASP...79..102A}. Based on the 2MASS photometry, \citet{masana2006} computed \teff{}\,=\,7772\,K. The calculated value of \teff{}  by different authors are 7500\,K \citep{joshi2009}, 7638\,K \citep{2011MNRAS.414..792B} and 7177\,K  \citep{2012MNRAS.427..343M}. Using V-K$_s$ colors,  \citet{2014MNRAS.C} derived \teff{} as 7340\,K. On the basis of the time-resolved photo-electric photometry, \citet{2006A&A...455..303J} found $\delta$-Sct type pulsation of period 60-min and predicted the presence of multi-periodic pulsations. Our high-resolution spectroscopic analysis shows that Ca is underabundant and the group of lanthanide elements are overabundant. The absence of magnetic field and  pattern of elemental abundances place this star in the category of mild Am star.
	 
Fig.~\ref{fig:scatter_comp} shows the inter-comparison of derive  parameters (e.g. \teff{}, \logg{}, \luminosity{}) with the literature values summarized in Table~\ref{table:final_results}. The upper panels of this figure present the direct comparison while the lower ones show the residuals. For \teff{}, our determinations are consistent within the quoted errors which is visible in the bottom panel of the left most plot of this figure. Statistical comparison of derived and literature \teff{} values shows a mean difference of $103\pm108$\,K. Similarly, for \logg{}, our analysis reveals difference of $0.19\pm0.38$. For \luminosity{} values, we compared our determinations with the those obtained through empirical relation given by \citet{catanzaro2015} that reflects a mean difference of $0.24\pm0.39$ for HD\,13079 and HD\,102480. For rest of stars the mean difference is about $0.04\pm0.10$.

Two stars of our sample namely HD\,113878 and HD\,118660 have been previously studied spectroscopically by \citet{2014MNRAS.C}. On comparing these published spectroscopic parameters with the present ones (marked as 10c and 1a respectively in Table~\ref{table:final_results}) we found that our \teff{} determinations for these two stars are higher by 100 K and 350 K, respectively and those are within the error limits of 400 K and 350 K, respectively. Similar agreement is valid for \logg{} where the difference for these two stars are 0.30 and 0.10 respectively. The small disagreements between these two sets could be due to fact that different approaches were used for the determination of these parameters.

	\section{Discussion and Conclusions}
	\label{section:conclusion}
	
We present the measurements of atmospheric parameters, chemical abundances and strength of longitudinal magnetic field in the sample stars where $\delta$ Sct type pulsations were discovered under the ongoing Nainital-Cape survey project. Such studies are particularly important as they are treated as a potential test bed to probe the stellar interior in the presence of various physical processes such as diffusion, magnetic field etc. 
	
	The primary goal of the present study was to measure the basic physical parameters of the CP stars and establish their chemical peculiarities. We applied photometric and spectroscopic techniques to derive the basic parameters and found that \teff{} and \logg{} determined from different methods are generally in agreement. Our analysis shows that the sample stars have effective temperature between 6720\,K and 7960\,K and  gravity in the range of 2.9 to 4.0. The stars in our sample have low-rotational velocity i.e. 38 to 108 km\,s$^{-1}$ producing a basic requirement to induce the diffusion process for producing the chemical peculiarities. Based on the derive parameters, we placed the sample stars in the H-R diagram and found that all are evolved from ZAMS and moving towards TAMS or slightly beyond it and within the uncertainties they are located inside the $\delta$ Sct instability strip. We compared the atmospheric parameters derived through different methods and found that within errors they generally agree to those complied from the literature.
	
	  The comprehensive atmospheric abundances of the sample stars were derived using the spectral synthesis method.  In HD\,13038, vanadium (V) is overabundant by factor of 1.71 while zinc (Zn) is underabundant by 1.04. HD\,13079 is a multi-periodic pulsating Am star where abundance of most of the elements are solar value, though cerium (Ce) is overabundant by 0.8. Scandium (Sc) is almost underabundant by 1.61 in HD\,25515. Two unusual pulsators HD\,98851 and HD\,102480 have similar chemical properties and abundance of majority of elements are of solar value. In HD\,113878, the group of lanthanide elements such as Ba, La, Ce etc are overabundant and Sc is underabundant.  Among our sample stars, HD\,113878 and HD\,118660 were also studied by \citet{2014MNRAS.C} in high-resolution spectroscopic mode and our findings are consistent to those obtained by these authors.  
 
	The secondary aim of our investigation was to establish the presence of magnetic field because the strength of magnetic field distinguish two classes of CP stars, Ap and Am stars. For this, we performed the spectropolarimetric analysis of the sample stars and found that none of the programme stars have significant magnetic field in the line of the sight of observer, an another evidence for their Am classification.
	
 We notice that one of sample star HD\,118660 shows presence of strontium (Sr) and neodymium (Nd) and overabundant by factor of 1.38 and 0.96 dex, respectively. Furthermore, HD\,118660 has the maximum rotational broadening among our sample (Table~\ref{table:final_results}). Due to these additional peculiar characteristics, we acquired time-resolved spectroscopic data of this star with NES mounted on 6.0-m telescope of SAO. The preliminary analysis revealed that line profile of some elements are significantly changing with time. In caontrary to the single mode pulsation reported by \citet{1999MNRAS.309..871M} in one of sample star HD\,13079, the data obtained on this star using WASP survey, \citet{smalley2011} detected five  pulsational frequencies. Therefore, in near future, we propose to monitor all the sample stars using time-series photomeric and spectroscopic means to search for and study multi-periodic pulsational and rotational variability for their asteroseismic studies.   
		
\section*{Acknowledgments}
	
The authors thank Dr. Barry Smalley for the constructive comments and suggestions that substantially improved the original version of the paper. The work presented in this paper is carried out under the Indo-Russian project INT/RFBR/P-118  and RFBR Grant No. $\textrm{12-02-92693-IND{\_}a}$ jointly supported by DST, India  and RAS, Russia. ES is thankful to the Russian Science Foundation (grant No. 14-50-00043) for financial support. KS thanks Neelam Panwar for assistantship through DST-INSPIRE faculty award. YBK thanks to CAS and NSFC for the support. The research was made with using of the SIMBAD database and VizieR catalogue access tool, operated at CDS, Strasbourg, France. This work has made use of data from the European Space Agency (ESA) mission {\it Gaia} (\url{http://www.cosmos.esa.int/gaia}), processed by the {\it Gaia} Data Processing and Analysis Consortium (DPAC, \url{http://www.cosmos.esa.int/web/gaia/dpac/consortium}). Funding for the DPAC has been provided by national institutions, in particular the institutions participating in the {\it Gaia} Multilateral Agreement. It has has made use of the VALD data base, operated at Uppsala University, the Institute of Astronomy RAS in Moscow, and the University of Vienna. 

\bibliographystyle{mnras}
\bibliography{ref}
\bsp
\label{lastpage}
\end{document}